\documentclass[preprint,3p,12pt,authoryear,aps]{revtex4}
\usepackage[pdftex]{graphicx}
\usepackage{dcolumn}% Align table columns on decimal point
\usepackage{bm}% bold math
\usepackage{wasysym}
\usepackage{amssymb}

 \usepackage[T1]{fontenc}

\usepackage{float}
\usepackage{hyperref}

\begin{document}

\title{Effect of electrolyte on the microstructure and yielding of aqueous dispersions of colloidal clay}

\author{Samim Ali}
 \email{samim@rri.res.in}
\affiliation{Raman Research Institute, C. V. Raman Avenue, Sadashivanagar, Bangalore 560080, India}
\author{Ranjini Bandyopadhyay}
\email{ranjini@rri.res.in}
\affiliation{Raman Research Institute, C. V. Raman Avenue, Sadashivanagar, Bangalore 560080, India}


\begin{abstract}

 Na-montmorillonite is a natural clay mineral and is available in abundance in nature. The aqueous dispersions of charged and anisotropic platelets of this mineral exhibit non-ergodic kinetically arrested states ranging from soft glassy phases dominated by interparticle repulsions to colloidal gels stabilized by salt induced attractive interactions.  When the salt concentration in the dispersing medium is varied systematically, viscoelasticity and yield stress of the dispersion show non-monotonic behavior at a critical salt concentration, thus signifying a morphological change in the dispersion microstructures.  We directly visualize the  microscopic structures of these kinetically arrested phases  using cryogenic scanning electron microscopy.  We observe the existence of honeycomb-like network morphologies for a wide range of salt concentrations. The transition  of the gel morphology, dominated by  overlapping coin (OC) and house of cards (HoC) associations of clay particles at low salt concentrations to a new network structure dominated by face-face coagulation of platelets,  is observed across the critical salt concentration. We further assess the stability of these gels under gravity using electroacoustics. This study, performed for  concentrated clay dispersions for a wide concentration range of externally added salt,  is useful in our understanding of many geophysical phenomena  that involve the salt induced aggregation of natural clay minerals.

\end{abstract}
\maketitle

%Valid PACS numbers may be entered using the \verb+\pacs{#1}+ command.
\section{Introduction}

\indent Aqueous dispersions of smectite clay minerals have been investigated extensively during the last decade to understand their rich phase behaviors \cite{Abend.2000.1, Jabbari.PRE.2008}, aging dynamics \cite{Ranjini.PRL.2004, Negi.JoR.2010, Debasish.SoftMatter.2014, tudisca.RSCAdv.2012, debasish.Langmuir.2015} and unusual flow properties \cite{fossum.EPJ.2012, Bailey.SoftMatter.2015}. The bulk properties of these dispersions originate from the complex self organization of charged anisotropic clay  platelets \cite{Sharon.NatMat.2007}, which leads to various phases such as gels, glasses \cite {ruzicka.SoftMatter.2011}, empty liquids, equilibrium gels \cite{ruzicka.NatureMaterial.2011} and nematic liquid crystals \cite{Paineau.LCR.2013}. These phases and their  bulk behaviors can easily be realized by tuning the colloidal interactions externally. This has led to widespread applications of these  clay minerals as rheological modifiers and stabilizers in paints \cite{faheem.MMT-A.2008}, well bore drilling \cite{Murray.2000.207}, cosmetics \cite{viseras.ACSc.2007}, pharmaceuticals \cite{viseras.ACSc.2007}, agrochemicals and nanocomposites \cite{Liu.2006.1}.

 \indent The charged anisotropic platelets of the Na-montmorillonite clay mineral used in this work are naturally occurring flexible nanosheets \cite{Bergaya.2013.1}. The platelets are highly polydisperse in their lateral sizes and shapes. The surfaces of these platelets have negative charges while the edges have pH dependent positive charges. In aqueous dispersions, the clay platelets are surrounded by hydrated Na$^{+}$ counterions that form anisotropic electric double layers (EDL) around the platelets. Due to high concentration of negative charges on their basal surfaces, there is a small amount of spill-over of the negative potential onto  the positively charged edges at pH$>$7 \cite{Secor.JCIS.1985, Yan.JPCB.1999, Zhou.EST.2012}. Due to the presence of the anisotropic EDL, the effective excluded volume of each platelet is much higher than its geometrical volume. The interaction potential between these platelets in dilute samples is usually represented by the DLVO theory \cite{Overbeek-book-1948} and depends on the mutual separation, the relative orientations and  the charge distribution on the platelets. Thus, the formation of kinetically arrested  phases such as soft glasses and gels in the dispersions can be controlled by changing clay mineral concentration, salt concentration and pH of the medium \cite{Abend.2000.1}. The rheological and stability properties of these disordered phases are directly related to the microscopic arrangements of the platelets in the dispersions. In the glassy phase, the interactions between platelets are dominated by   screened Coulomb repulsions and the viscoelasticity and yielding behaviors originate from the caging of each platelet by its neighbors \cite{Tanaka.PRE.2004}. The gel phase, on the other hand, is a volume spanning network structure in which two platelets are connected by an attractive bond, with the strengths of the bonds determining their viscoelasticity and yielding behavior. \\

\indent The nature of  platelet association for attractive bond formation, and therefore the origin of the observed rheological behavior of Na-montmorillonite gels, have been widely debated in the literature \cite{norrish.1954, Bowles.Science.1968, Olphen_book, Michot.Langmuir.2013}. Based on the DLVO theory and rheological measurements in dispersions of low platelet concentrations, it was predicted that at low salt concentration and in a dispersing medium of pH$<7$,   attractive bond formation occurs through the association of the negative faces of the platelets with their positive edges. On the other hand,  bond formation through edge-face, face-face and edge-edge interactions are predicted at varying salt concentrations in a dispersing medium of pH$>7$ \cite{duran.JCIS.2000, Lagaly.2003.1}. The coagulation process becomes more complicated with increasing platelet concentration   in the presence of salt and a  house of cards (HoC) structure is frequently invoked to  explain the nature of gel networks \cite{broughton.JPC.1935, vanOlphen.JCS.1964, Au.CSA.2013, kimura.RheoActa.2011, Laxton.JCIS.2006, dijkstra.PRL.1995}. Surprisingly, the experimental evidence of such predicted microstructures is very limited.  There have been some studies to visualize the underlying microscopic structures using scanning electron microscopy (SEM) \cite{wierzchos.CCMin.1992} and  transmission X-ray microscopy (TXM) \cite{marek.Langmuir.2008, Michot.Langmuir.2013} for different conditions of sample preparation. These studies confirm the existence of either edge-edge or face-face microscopic configurations of platelets in the presence of salt. However, these studies do not systematically investigate the variation of the microscopic structures, and their influence on the strength and stability of clay gels, with changing salt concentration.

	\indent In this article, we address this issue by studying the rheology, stability  properties and associated microstructures of 5\% w/v Na-montmorillonite dispersions at their natural pH values after systematically varying the externally added salt concentration from 10 mM to 800 mM. We find that  the viscoelastic moduli and the yield stress  of arrested phases (gels) in the dispersion increase upto a peak value at a critical salt concentration  and subsequently decrease due to the progressive increase in salt induced interparticle attractive interactions.  The microscopic association of the platelets in gels with varying salt concentration  is directly visualized  using cryogenic scanning electron microscope (cryo-SEM). This  shows  a transition  of the gel morphology, dominated by  overlapping coin (OC) and house of cards (HoC) associations of clay particles to a new network structure dominated by face-face coagulation of platelets, across the critical salt concentration. The variation of the strength of the gels estimated from the rheological measurements is then interpreted in terms of the observed microstructures and changes in gel morphology.  The influence of the morphology on the stability of the gels under gravity is further assessed using electroacoustics.  	
	
\indent 	Besides their obvious rheological importance \cite{Bergaya.2013.1}, clay colloids in aqueous dispersions have been studied extensively for their aging properties \cite{tanaka.PhysRevE.2005, angelini.Naturecom.2014}. The structure, dynamics and rheology of clay in water with and without salt have been widely debated \cite{fossum.EPJ.2012, Olphen_book, Michot.Langmuir.2013}. Our present results, obtained for clay dispersions with a wide range of salt concentrations, will be useful to understand and predict  many geophysical phenomena such as land slides, and the formation of river deltas and quicksand \cite{Thill.CSR.2001, khaldoun.Nature.2005} that are directly or indirectly related to the salt induced association of natural clay colloids.

\section{Experimental section}
\subsection{Material Structure}
  
	\indent We use Na-montmorillonite of CEC value 145 meq/100g procured from Nanocor Inc \cite{nanocor-mmt}. A unit layer of this mineral is comprised of  2:1 layered phylosilicate \citep{VanOlphen.1962.1,Zheng.2011.80}. The general formula is Na$^{+}_{x}$[(Al$_{2-y}$Mg$_{y}$)Si$_{4}$O$_{10}$(OH).nH$_{2}$O]$^{-}$ \cite{Bergaya.2013.1}. Each unit layer, also known as a platelet, consists of an aluminum octahedral sheet sandwiched between two tetrahedral silica sheets. The thickness of a platelet is around 1 nm. The lateral size of these platelets may vary from tens of nanometers to a few micrometers.  In dry form, several platelets form a stack, known as a tactoid, with intercalated   Na$^{+}$ counterions. In aqueous dispersions,  the Na$^{+}$ ions get hydrated due to the absorption of water molecules in the intratactoid spaces.   As a result, tactoids slowly swell and   exfoliate, producing laminar flexible platelets with electric double layers (EDLs) on their surfaces \cite{ramsay.1990.1, Cadene.2005.1, ali.samim.2013.1, ali.ACSs.2015}. The exfoliated platelets are highly irregular in shape and size (shown in a representative SEM micrograph in Fig. S1 of the ESI\ddag).  An average lateral size of 450 nm is calculated from the SEM micrographs (Fig. S2, ESI\ddag).

 \subsection{ Sample preparation}
 \indent The Na-montmorillonite powder is baked for 24 hours in an oven kept at a temperature of $120^{\circ}$C to remove moisture. A stock dispersion of 8\% w/v is then prepared by dispersing the dry powder in highly deionized Milli-Q water under vigorous stirring conditions using a magnetic stirrer.  The dispersion is homogenized by stirring it for three hours and  then stored in a sealed polypropylene bottle for seven days. The stock dispersion is next used to prepare 5\% w/v clay dispersions with different ionic strengths by adding predetermined quantities of NaCl solutions. The mixture of clay and salt solution is next stirred for three hours using a magnetic stirrer. The resultant dispersions are   kept in vacuum   for two minutes to remove air bubbles trapped in the viscous medium. Samples with different salt concentrations, $C_{s}$, are then stored for four days   in sealed glass vials before using them for    rheological measurements, cryo-SEM imaging and electroacoustic measurements. The pH of these dispersions are maintained at their natural values. The pH measurements are done using a CyberScan Eutech electrode (Model-ECFG7252001B) at a temperature of $25^{\circ}$C. The addition of salt leads to a slight decrease in the pH value of the dispersion and has been shown in Fig. S3 of the ESI\ddag. It is seen that the pH of the dispersion always remains above 8.8 in the salt concentration range investigated here.

\subsection{Experimental setups and measurements} \label{methods}

	\noindent  \textbf{Cryogenic scanning electron microscopy (cryo-SEM):}
	
	%Show images of the cryo-SEM sample holder ans samples in supplementary info.
	\indent  For cryo-SEM characterization, shear melted samples with different salt concentrations are loaded by a syringe in capillary tubes (procured from Hampton Research, USA) of bore size  1 mm. The ends of the capillaries are then quickly sealed. Samples are then kept in an undisturbed condition for 48 hours. A home made sample holder is used for holding the sample capillaries. The samples are then vitrified using liquid nitrogen slush of temperature $-200^{\circ}$C. The vitrified samples are then fractured, sublimated for 12 mins at a temperature $-90^{\circ}$C and coated with a thin layer of platinum  at a temperature $-150^{\circ}$C in  vacuum  using a cryotransfer system (PP3000T  from Quorum Technologies).  The imaging of these samples is then performed using a field effect scanning electron microscope (FESEM) from Carl Zeiss at an electron beam strength of 2 KeV.

 \noindent \textbf{Rheology:}
	
	\indent  Rheological measurements are perform by an Anton Paar MCR 501 rheometer working in a stress-controlled oscillatory mode. For each rheological experiment, a  couette geometry is filled carefully with 4.7 ml of sample using a syringe. The filling process partially rejuvenates the sample.  The free surface of the sample is covered with a thin layer of silicon oil of viscosity   5 cSt to prevent   evaporation of water. A well defined experimental protocol, as shown in the inset of Fig. S4 of the ESI\ddag, is used for all the measurements. After filling the measurement geometry, the samples are shear rejuvenated by applying a large oscillatory stress of amplitude 50 Pa with an angular frequency of 6 rad/s. The application of this high shear stress liquifies the samples, leading to zero elastic modulus $G'$ and very small values of viscous modulus $G''$  (Fig. S4 of the ESI\ddag). The samples are  left to evolve for three hours after cessation of the shear melting process at age $t_{w}=0$. During this period, the dispersions undergo a spontaneous phase transition from a liquid-like state to a kinetically arrested state, with the viscoelastic moduli evolving continuously with age $t_{w}$ (Fig. S5, ESI\ddag).  After $t_{w}=3$ hours, a strain amplitude sweep test is carried out by varying strain amplitude, $\gamma$, in the range of $0.1 - 100$\% at a constant angular frequency of 6 rad/s (inset of Fig. S4 of the ESI\ddag). All experiments reported here are carried out at a temperature of $25^{\circ}$C.

 \vspace{6mm} 
 \noindent  \textbf{Electroacoustics measurements:}
	
	\indent A cylindrical   electroacoustic probe supplied by Dispersion Technology Inc is used to monitor the stability of clay   dispersions under gravity. Details of the probe can be found in \cite{Dukhin-book-2nd}. The ultrasound transducer in the probe is co-axially placed and then insulated from the outer metal body. The sedimentation setup using this probe is shown in Fig. S6 of the ESI\ddag. In this setup, the front flat surface of the probe touches the the top surface of the dispersion column. A very thin layer of silicon oil of viscosity  5 cSt (at 20$^{o}$C) is placed between the probe and the dispersion. This keeps the sample from sticking to the flat surface of the probe. When the transducer launches a low power (10 mW)  nondestructive plane ultrasound wave of frequency 3 MHz along the height of the sample, it induces  small oscillating dipole moments in the EDLs of the clay platelets suspended in the dispersion. The electric field,  generated from these induced dipoles, in turn, induces a current known as the colloidal vibration current ($I_{CVI}$) in the receiving transducer circuit of the probe. If the dispersion contains salt, the transducer simultaneously detects an ionic vibration current ($I_{IVI}$) which arises due to the relative motion between the two ion species of the salt. In this case, the total electroacoustic signal ($I_{TVI}$) is a vector sum of $I_{CVI}$ and $I_{IVI}$.     The value of  $I_{TVI}$ measured in this setup is very sensitive to the distance of the clay particles comprising the gel from the measuring transducer \cite{ali.ACSs.2015}. Thus any shrinkage or settling of the gels will lead to a change in the measured values of $I_{TVI}$.

\begin{figure} [H]
\begin{center}
\includegraphics[width=5in]{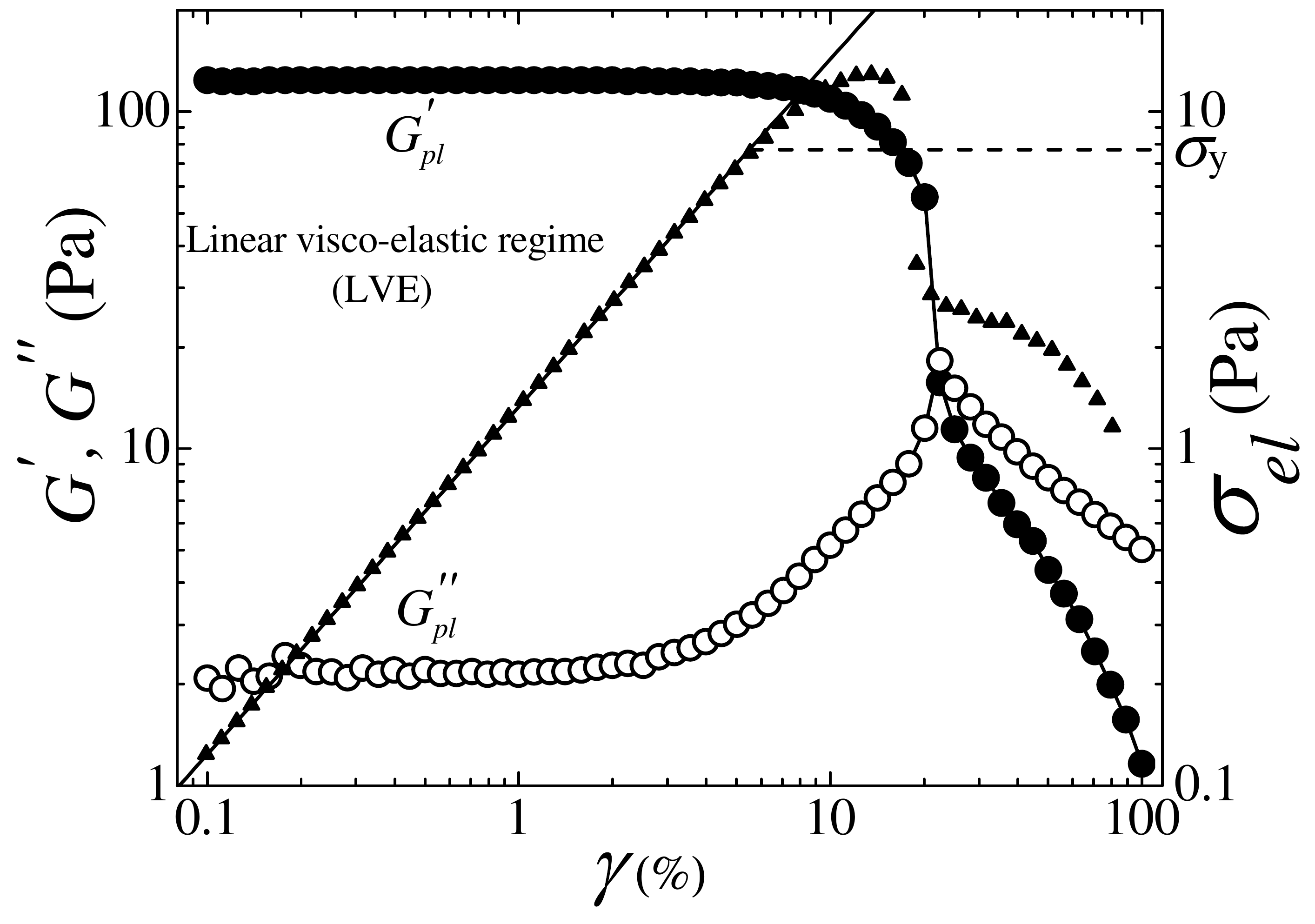}
\caption{ Variation of elastic modulus $G'$ ($\CIRCLE$), viscous modulus $G''$ ($\Circle$) and elastic stress, $\sigma_{el}=G'\gamma$ ($\blacktriangle$) with strain amplitude, $\gamma$, for 5\% w/v Na-montmorillonite with 20 mM added salt at $t_{w}=3$ hrs. The solid line is a linear fit to the $\sigma_{el}$ vs $\gamma$ data.   The  value of $\sigma_{y}$ is shown by the horizontal dashed line.\\}
\label{Fig:as-20mM}
\end{center}
\end{figure}

\section{Results and discussion}
 \indent The bulk mechanical behavior of Na-montmorillonite dispersions, with varying salt concentration $C_{s}$, is measured by  performing strain amplitude sweep tests. \autoref{Fig:as-20mM} shows  representative data of a strain amplitude sweep experiment  at an angular frequency of 6 rad/s  for a 5\% w/v clay dispersion with $C_{s}=20$ mM at $t_{w}=3$ hrs. At small values of applied strain ($\gamma$), i.e., in the linear viscoelastic (LVE) regime, $G'>G''$, with both the moduli being independent of strain amplitudes. The plateau values of $G'$ and $G''$ are designated by $G'_{pl}$ and $G''_{pl}$ respectively. On further increase in $\gamma$, the sample starts yielding due to the irreversible rearrangement of Na-montmorillonite platelets. In this nonlinear regime, $G'$ decreases monotonically while $G''$ reaches a peak at the point of crossover between $G'$ and $G''$.  Finally, at very high strains, the sample exhibits fluid-like behavior which is indicated by $G''>G'$. Similar behavior of the stress moduli under high applied strains in the nonlinear regime was also observed previously in simulation and experimental studies \cite{miyazaki.EPL.2006, laurati.JOR.2011}.The dynamic yield stress, $\sigma_{y}$ (indicated by horizontal dashed line in \autoref{Fig:as-20mM}), is calculated from the stain amplitude sweep data following the method described by Laurati \textit{et al.} \cite{laurati.JOR.2011}. In this method, the elastic stress $\sigma_{el}=G'\gamma$ ($\blacktriangle$ in \autoref{Fig:as-20mM}) is plotted versus $\gamma$, which helps  to separate the contribution of the viscous stress from the total stress. At low $\gamma$ values, the elastic stress vs strain data is fitted to $\sigma_{el}=G'\gamma$. The value of $\sigma_{y}$ is defined as the magnitude of $\sigma_{el}$ at which the measured value of $\sigma_{el}$ starts deviating from the theoretically calculated value. For the sample in \autoref{Fig:as-20mM}, dynamical  yield stress $\sigma_{y}=6.8$ Pa is calculated.

\begin{figure}
\begin{center}
\includegraphics[width=5in]{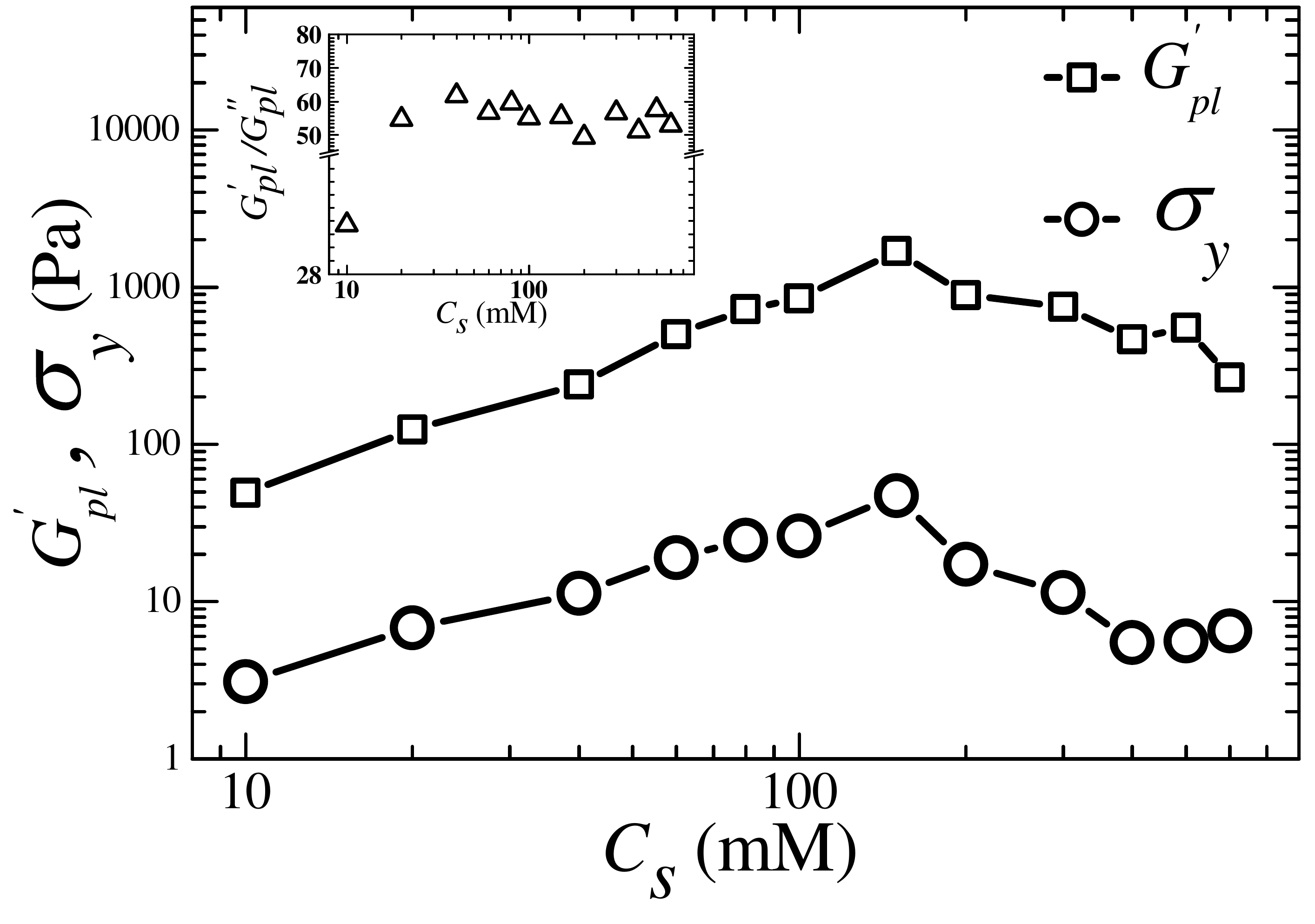}
\caption{ Variation of plateau value of elastic modulus $G'_{pl}$ ($\square$) and dynamic yield stress $\sigma_{y}$ ($\Circle$) with increasing salt concentration $C_{s}$. The inset shows the change in ratio of viscoelastic moduli $G'_{pl}/G''_{pl}$ ($\triangle$), measured in the  linear viscoelastic regime, versus $C_{s}$.\\}
\label{Fig:AS-para}
\end{center}
\end{figure}

	\indent The variations of $G'_{pl}$ and $\sigma_{y}$ with varying salt concentration $C_{s}$ are shown in   \autoref{Fig:AS-para}.  Both the quantities  increase monotonically with $C_{s}$ upto  $C_{s}\approx150$ mM before they decrease with further addition of salt. The inset of \autoref{Fig:AS-para} shows that  $G'_{pl}$ is approximately sixty times higher than $G''_{pl}$ for samples with $C_{s}>10$ mM. This indicates that the dispersions are essentially elastic in the linear viscoelastic regime, even though the   strength of the underlying microstructures of the dispersions  decreases for $C_{s}>150$ mM. The observations in  \autoref{Fig:AS-para} therefore indicate a transition in the sample  morphology at $C_{s}=150$ mM.

\begin{figure} 
\begin{center}
\includegraphics[width=6in]{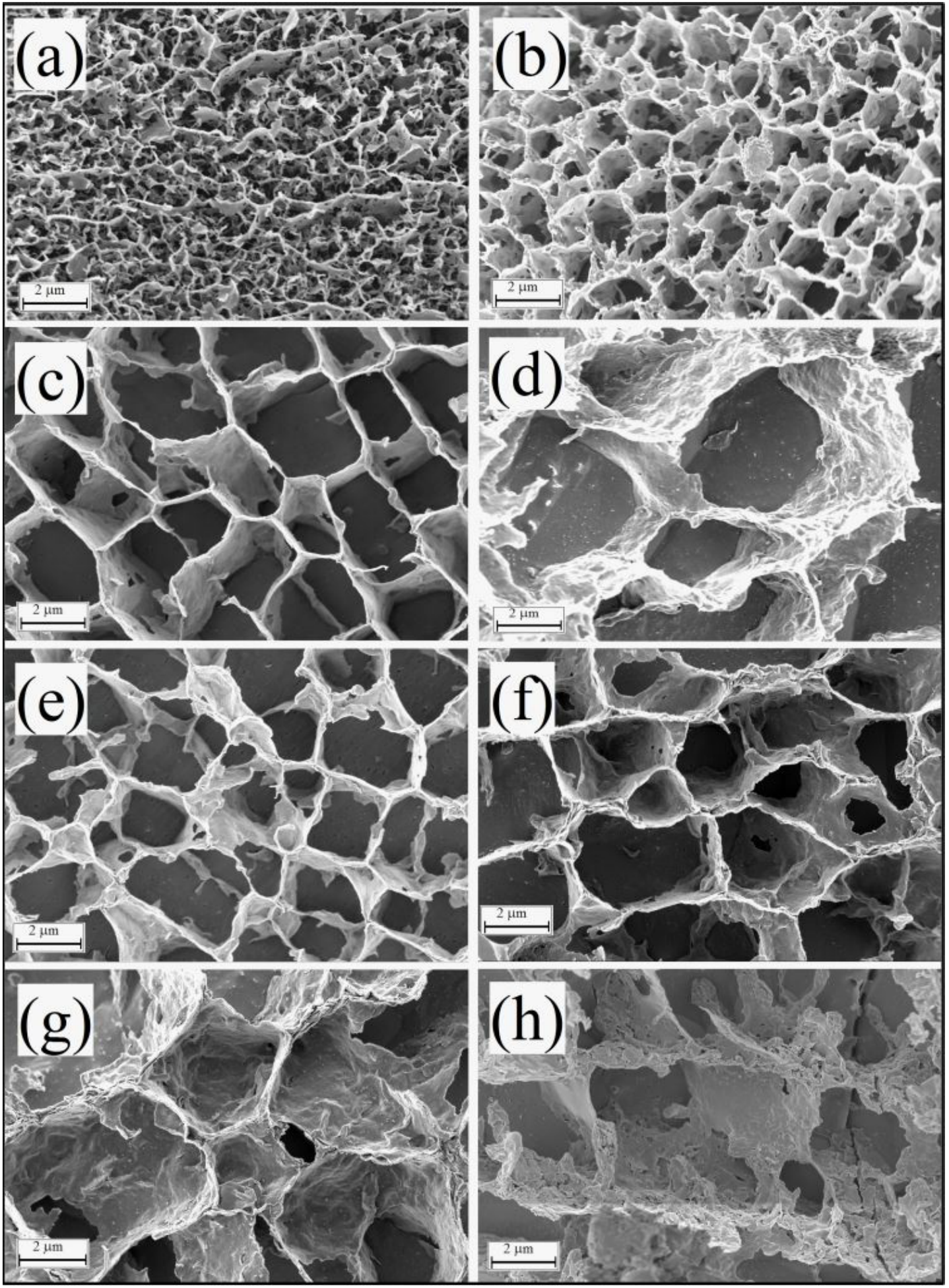}
\caption{ Representative micrographs obtained using cryo-SEM for 5\% w/v Na-montmorillonite dispersions with $C_{s}=$ 10 mM (a), 20 mM (b), 50 mM (c), 100 mM (d), 150 mM (e), 200 mM (f), 300 mM (g) and 500 mM (h). The scale bars represent 2 $\mu$m.\\}
\label{Fig:Cryo_SEM-images}
\end{center}
\end{figure}

\begin{figure} 
\begin{center}
\includegraphics[width=4in]{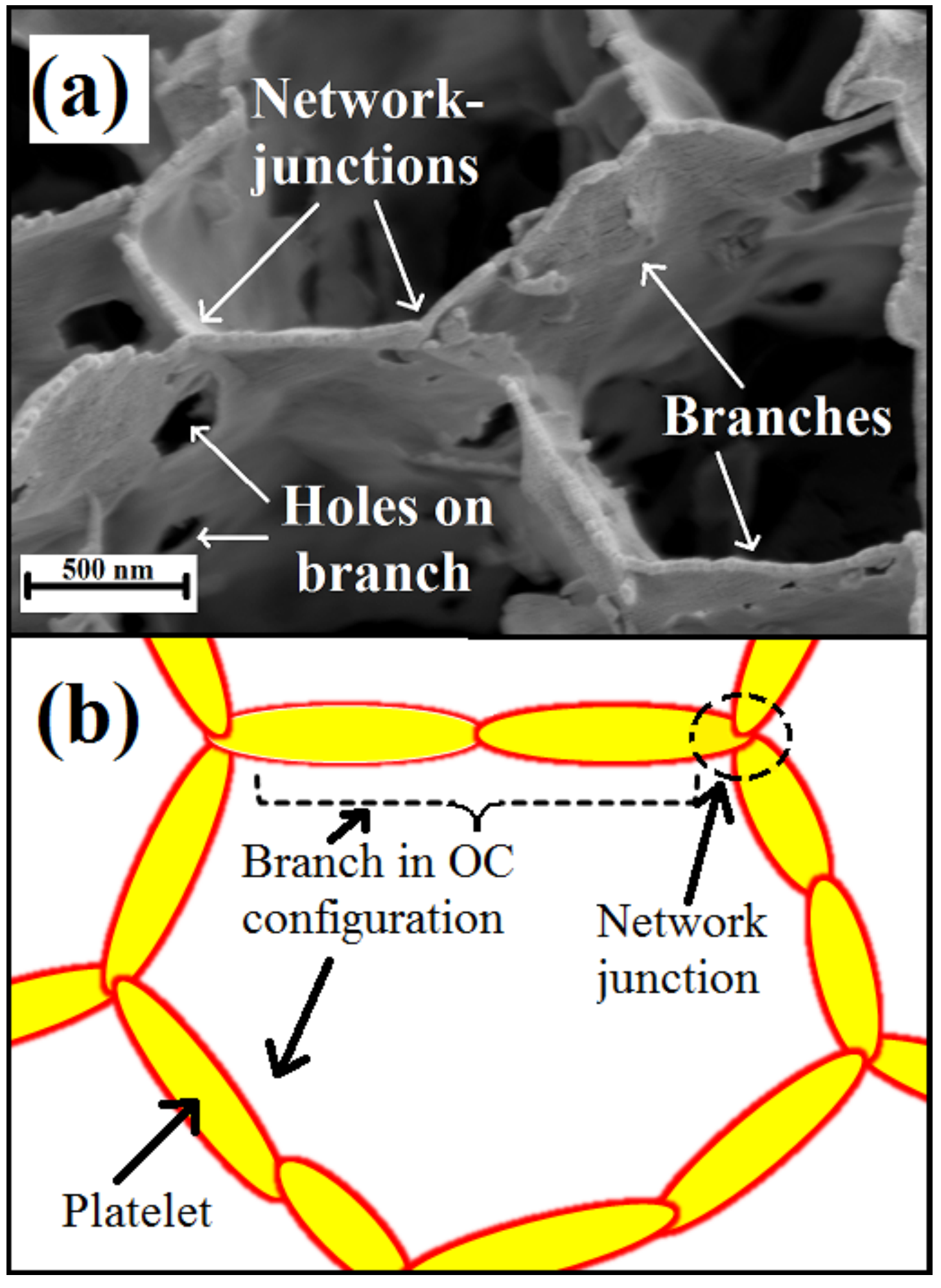}
\caption{ (a) Magnified view of a representative micrograph obtained using cryo-SEM for 5\% w/v Na-montmorillonite dispersion with $C_{s}=$ 20 mM. (b) Schematic depiction of the microscopic arrangement of circular platelets showing overlapping coins (OC) in dispersion in the presence of salt. The red color on the edges indicates positive charges  and the yellow color on the basal surface indicates screened negative charges.\\}
\label{Fig:SEM-20mM-schematic}
\end{center}
\end{figure}

\indent We next investigate the salt induced morphological changes of Na-montmorillonite gels using cryo-SEM. \autoref{Fig:Cryo_SEM-images} shows representative cryo-SEM micrographs of 5\% w/v Na-montmorillonite gels  with $C_{s}$ varying in the range 10-500 mM. Honeycomb-like three dimensional network structures, with a systematic change in  morphology and sizes of the polydisperse pores (voids left after sublimating the water molecules during the cryo-SEM sample preparation step), can be observed in all these samples. The  branches of the gel networks observed here are thicker  than the thickness  of a single platelet  ($\sim$1 nm) due to the presence of   vitrified water on their surfaces.  A close inspection of the honeycomb structures formed by 5\% w/v Na-montmorillonite and 20 mM salt (\autoref{Fig:Cryo_SEM-images} (b)) reveals that the average length of the branches is larger than the average lateral size (450 nm) of the Na-montmorillonite platelets. A magnified image of this sample is provided in  \autoref{Fig:SEM-20mM-schematic} (a) which clearly shows that    many of the branches have holes on their surfaces. This  indicates that the platelets on each branch   are connected in overlapping coin (OC) configurations, as predicted by Jonsson \textit{et. al.}  using Monte Carlo simulations  in a system of clay platelets at low salt concentration \cite{jonsson.Langmuir.2008, delhorme.SoftMatter.2012}. In the OC configuration, the positive edge of a platelet attaches to the negative basal surface  near   the edge of another platelet in a parallel fashion, thereby forming longer sheets (branches of the network) through attractive bonds  (\autoref{Fig:SEM-20mM-schematic}(b)).  The positive edges of two such sheets (the branches) attach to the   negative faces of a third  sheet comprising platelets which are also in OC configurations. Such attachments  lead to the formation of an attractive network-junction   of three branches as indicated in \autoref{Fig:SEM-20mM-schematic}.  We note here that such honeycomb-like network formation is not very dominant in the case of the sample with 10 mM salt (a magnified image is shown in Fig. S7 of the ESI\ddag) due to the presence of high face-face repulsions.

	\begin{figure} [!t]
\begin{center}
\includegraphics[width=4in]{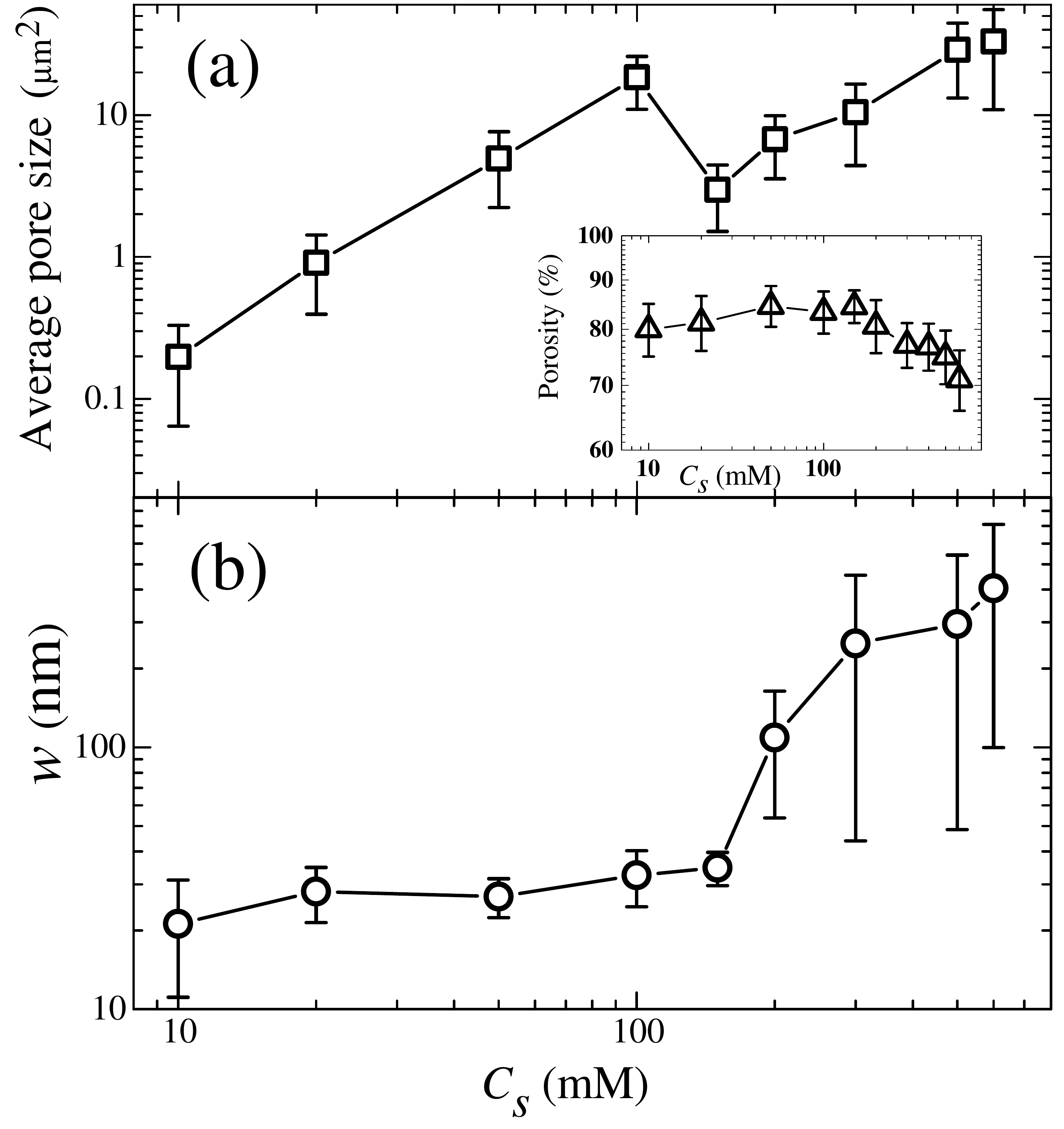}
\caption{ (a) Plot of average pore size ($\square$) of the gel network as a function of salt concentration $C_{s}$. Inset shows the variation of   porosity ($\triangle$) with $C_{s}$. (b) Plot of network-branch thickness, $w$, ($\Circle$) as a function of $C_{s}$. The image analysis is performed using ImageJ.  \\}
\label{Fig:poresize-thickness-salt}
\end{center}
\end{figure}

\begin{figure} 
\begin{center}
\includegraphics[width=2in]{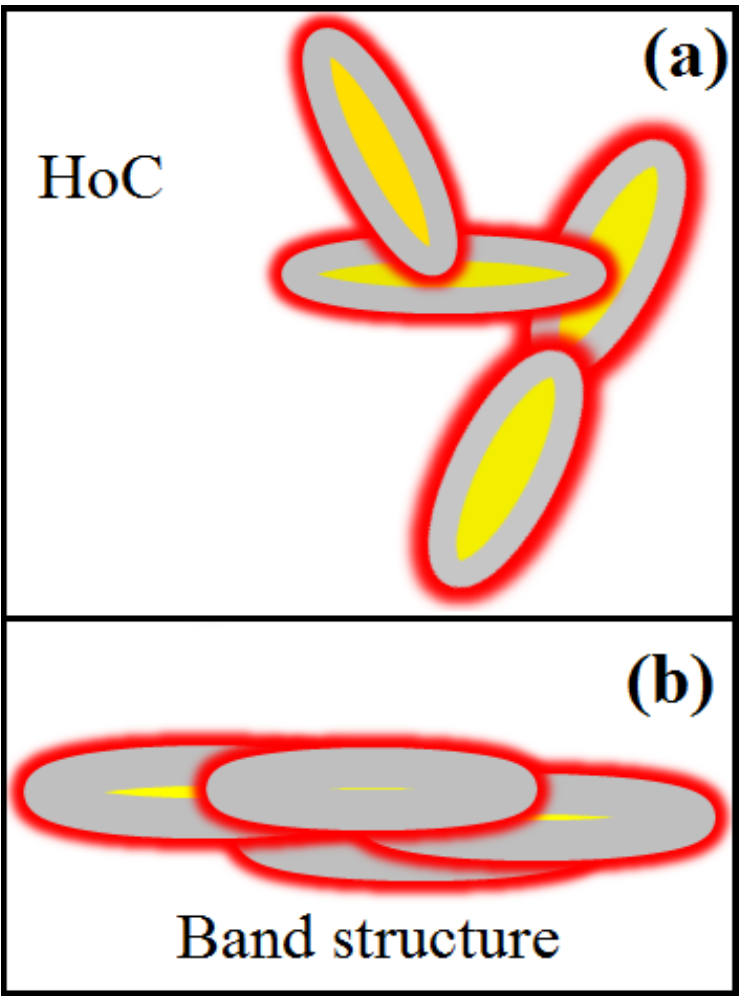}
\caption{ Schematic depiction of (a) house-of-card (HoC) arrangement of circular platelets and (b) face-face aggregation of platelets leading to band-type structures in the dispersion in the presence of salt. The red color on the edges indicates positive charges  and the yellow color on the basal surface indicates screened negative charges. The grey color indicates complete charge screening.\\}
\label{Fig:Cartoon-2}
\end{center}
\end{figure}

\indent We use  the cryo-SEM micrographs to quantify the pore (void) size distributions, porosity  and branch thicknesses  of all the gels studied in this work. Porosity  is defined as a ratio of the total void area to the total area of  the 2D projection of the  gel structure. The image analysis tools used here and the details of the calculation of the porosity and network branch thickness are discussed in Section A, and in Figs. S8 and S9 of the ESI\ddag. It is to be noted that the pore sizes measured here are smaller than the actual sizes due to the presence of   vitrified water on the network. However, since the sublimation time (12 min)  after fracturing  the vitrified samples is same for all the dispersions,  an equal  sublimation-depth is expected for all the samples studied using cryo-SEM. When $C_{s}$ is increased systematically upto 100 mM, the average pore size ($\square$ in \autoref{Fig:poresize-thickness-salt}(a)) increases, while  the porosity ($\triangle$ in the inset of \autoref{Fig:poresize-thickness-salt}(a)) and branch thickness $w$ ($\Circle$ in \autoref{Fig:poresize-thickness-salt}(b)) of the gels remain almost unchanged. It can therefore be concluded that for $C_{s}\le 100$ mM, the participation of platelets in OC configurations increases. The number of network-junctions simultaneously decreases, while the lengths of the individual branches of the network increases with $C_{s}$. This was clearly observed in \autoref{Fig:Cryo_SEM-images}(a), (b), (c) and (d).  Interestingly, at $C_{s}=150$ mM, the average pore size ($\square$ in \autoref{Fig:poresize-thickness-salt}(a))  decreases while the porosity ($\triangle$ in inset of \autoref{Fig:poresize-thickness-salt}(a)) and  $w$ ($\Circle$ in \autoref{Fig:poresize-thickness-salt}(b))   remain unchanged.  This reduction in pore size at $C_{s}=150$ mM (\autoref{Fig:Cryo_SEM-images} (e)) arises due to the participation of a substantial number of platelets in  house-of-cards (HoC) configurations, apart from the OC configurations discussed earlier. In such HoC configurations,  a positive edge of a platelet attaches attractively to the central negative part of the basal surface of another platelet (\autoref{Fig:Cartoon-2}(a)).    The coexistence of  HoC and OC configurations  can be seen in a magnified micrograph in  Fig. S10 of the ESI\ddag  for a sample with $C_{s}=150$ mM. Thus, at this salt concentration,  the lengths of the network branches decrease and number of network junctions increases due to the participation of a substantial number of platelets in HoC configuration. This leads  to the reduction in pore size at $C_{s}=150$ mM as observed in \autoref{Fig:poresize-thickness-salt}(a). It is to be noted that the  coexistence of such configurations was   predicted earlier in simulations of clay dispersions at high salt concentrations \cite{delhorme.SoftMatter.2012}.

  Since the addition of salt in the dispersing medium leads to a decrease in the Debye screening length, the spillover of negative potential from the basal surface onto the positive edges of platelets decreases with increase in salt concentration \cite{Secor.JCIS.1985, Yan.JPCB.1999, Zhou.EST.2012}. The effective potentials on the edges of the platelets are expected to be positive at $C_{s}=10$ mM. This leads to the formation of a gel network at this salt concentration as seen in  \autoref{Fig:Cryo_SEM-images}(a). With increasing $C_{s}$, the magnitude of the effective positive potentials on the platelet edges increases until the long-range effects of the negative charges of the basal surface becomes negligible due to screening.  As a result, the number of attractive bonds and their  strengths  in OC and HoC configurations  increase with $C_{s}$ (\autoref{Fig:Cryo_SEM-images}(a), (b), (c), (d) and (e)). This contributes to the growing $G'_{pl}$    and $\sigma_{y}$ of the gels upto $C_{s}=150$ mM (\autoref{Fig:AS-para}).

\indent At $C_{s}>150$ mM, the average pore size ($\square$ in \autoref{Fig:poresize-thickness-salt}(a)) and   $w$ ($\Circle$ in \autoref{Fig:poresize-thickness-salt}(b)) increase   with increasing $C_{s}$. This was clearly observed in \autoref{Fig:Cryo_SEM-images}(f), (g) and (h). The total void space simultaneously  deceases slightly ($\triangle$ in the inset of \autoref{Fig:poresize-thickness-salt}(a)). These features, and a close inspection of the micrographs (a representative magnified image of the sample \autoref{Fig:Cryo_SEM-images} (f) is shown in Fig. S11, ESI\ddag), reveal that  due to the considerable screening of negative charges on the basal surfaces, the face-face  aggregation process due to van der Waals attractions becomes   dominant  under very high salt conditions \cite{jonsson.Langmuir.2008}. 	Such parallel aggregation of platelets for low clay and high salt concentrations was also observed indirectly using ultrasound attenuation spectroscopy in our earlier study \cite{ali.ACSs.2015}. The face-face aggregation occurs very randomly and leads to elongated and thicker branches known as band structures (\autoref{Fig:Cartoon-2}(b)) \cite{weiss.Naturforsch.1961, permien.ACSc.1994, Luckham.ACIS.1999}. These band-type branches further connect at their ends and eventually form  a  kinetically arrested network characterized by a honeycomb structure. This can be seen in \autoref{Fig:Cryo_SEM-images}(e),(f),(g).  Incomplete network formation is also observed at very high salt concentrations, e.g. at $C_{s}=$ 500 mM (\autoref{Fig:Cryo_SEM-images}(h)) and 600 mM (Fig. S12, ESI\ddag) due to the strong face-face   aggregations of a considerable fraction of clay platelets.

 \begin{figure}
\begin{center}
\includegraphics[width=5in]{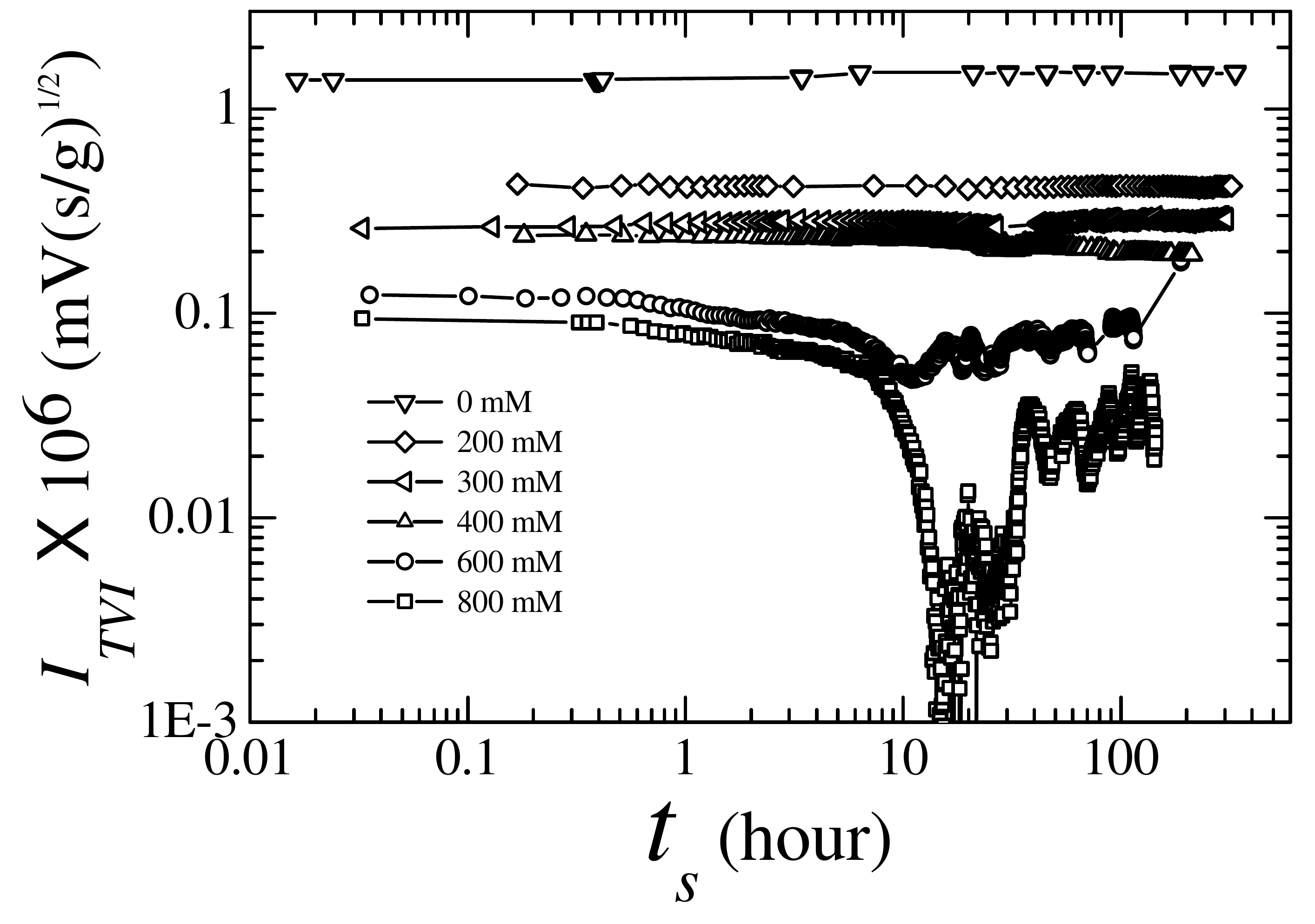}
\caption{ Evolution of total vibration current ($I_{TVI}$), normalized by the ultrasound pressure gradient, in a sedimentation setup (Fig. S6, ESI\ddag) with  observation time $t_{s}$. Here, $t_{s}=0$ is defined as the time when the stirring of the sample is stopped inside the sedimentation setup. The samples are  5\% w/v Na-montmorillonite with $C_{s}$= 0 mM($\triangledown$), 200 mM ($\diamond$), 300 mM ($\triangleleft$), 400 mM ($\triangle$), 600 mM ($\Circle$) and 800 mM ($\square$).\\}
\label{Fig:TVI-salt}
\end{center}
\end{figure}

 \indent The observed decrease in $G'_{pl}$  and $\sigma_{y}$ in \autoref{Fig:AS-para} when $C_{s}$ is increased beyond 150 mM can be attributed to the considerable increase in the pore sizes and   thicknesses of the network branches which is observed in \autoref{Fig:poresize-thickness-salt}. In this salt concentration regime, the excluded volume due to the basal charge repulsions decreases substantially with increasing $C_{s}$.  In addition, the strength of the attraction between the branches of the network also increases. As a result, applied strains promote irreversible aggregation of the thick branches. This process can lead to a higher pore size, larger branch thickness and decreasing network connectivity with increases $C_{s}$, leading to the observed decrease in $G'_{pl}$  and $\sigma_{y}$. Such coagulation of   network branches under stress can lead to the collapse of the gel network, with the yielding happening at lower $\gamma$ values as the network branches become increasingly thicker.\\

  We verify  the stability  of the gels under gravity by measuring  the  total vibration current ($I_{TVI}$) using a sedimentation setup described in  \autoref{methods}. \autoref{Fig:TVI-salt} shows the variation of  $I_{TVI}$ with observation time $t_{s}$,  measured for 5\% w/v Na-montmorillonite gels  for a $C_{s}$ range between 0 mM and 800 mM.  The dispersion with no added salt ($\triangledown$ in \autoref{Fig:TVI-salt}) is highly stable under gravity due to  the kinetic arrest of the constituent clay platelets.  Surprisingly, we find no change in $I_{TVI}$ values measured over two  weeks for samples with $C_{s}\le 300$ mM ($\diamond$  and $\lhd$ in \autoref{Fig:TVI-salt}). This indicates that the gels observed in \autoref{Fig:Cryo_SEM-images}(a)-\autoref{Fig:Cryo_SEM-images}(g) are highly stable under gravity. The sample with $C_{s}=400$ mM ($\triangle$ in \autoref{Fig:TVI-salt}) exhibits a small decrease in $I_{TVI}$   after $t_{s}=20$ hours. The dispersions with $C_{s}>400$ mM ($\Circle$ and $\square$ in \autoref{Fig:TVI-salt}) show irregular oscillations with observation time $t_{s}$ due to the intermittent collapse of the gel structures. As a result of the large pore sizes and  thick network branches of these samples (\autoref{Fig:Cryo_SEM-images} (h) and \autoref{Fig:poresize-thickness-salt}), the weight of the branches exceeds the local yield stress of the network, resulting in the observed gel collapse.

\section{Conclusions}

%Such transition if expected to be rather smooth.
\indent In this article, we present our results on  the influence of NaCl on the  microstructures, viscoelasticity, yielding and stability of  Na-montmorillonite gels at a clay concentration, 5\% w/v, at which the dispersion is expected to exhibit a glassy phase in the absence of salt \cite{ali.ACSs.2015}. Microscopic observations using cryogenic scanning electronic microscopy (cryo-SEM) reveal that at low salt concentrations, the clay platelets form   longer sheet structures (the network branches) through attractive overlapping coin (OC) configurations  predicted in a simulation study recently \cite{jonsson.Langmuir.2008}. These sheets (branches)  join at their ends to form  crosslinked ribbons (network-junctions), giving rise to   honeycomb-like network structures (\autoref{Fig:Cryo_SEM-images} (a), (b), (c)).  Interestingly, we find that platelet participation in the OC configurations in the network-branch increases with increase in salt concentration in the range of 10-100 mM. This leads to higher pore sizes  without any change in the   branch thickness $w$ of the gels (\autoref{Fig:poresize-thickness-salt}). With     further increase of $C_{s}$ upto 150 mM,   the thickness of the network branches   remains unchanged but the pore size of the gels  decreases with $C_{s}$ due to the participation of a substantial number of  clay platelets in the HoC configurations, besides the usual OC configurations (\autoref{Fig:Cryo_SEM-images}(d)). Such   a coexistence of OC and HoC at higher salt concentration was  also predicted in an earlier simulation  study \cite{delhorme.SoftMatter.2012}. Our rheological measurements (\autoref{Fig:AS-para}) further suggest that strength of the attractive bonds due to such platelet configurations increases with increasing   $C_{s}$ upto a value of 150 mM.

\indent At $C_{s}>150$ mM, the van der Waals attraction  between platelets becomes dominant due to the high screening of the  basal negative charges by the Na$^{+}$ ions. Under these conditions, the platelets  coagulate in a face-to-face orientation randomly, leading to the formation of elongated structures known as `band-type structures' that have been predicted in the literature (\cite{Luckham.ACIS.1999, weiss.Naturforsch.1961}). These elongated bands further connect at their ends, with the emergence of  kinetically arrested  honeycomb structures(\autoref{Fig:Cryo_SEM-images}(e), (f), (g)). The  pore sizes and branch thicknesses $w$ of the gel networks increase with $C_{s}$ due to the increase in the  face-face bond formation of the platelets (\autoref{Fig:poresize-thickness-salt}). The strength of the gels decreases with $C_{s}$  (\autoref{Fig:AS-para}) in this salt concentration regime due to  a decrease in the repulsive excluded volume and a simultaneous  increase in the attraction between the branches of gel network. This leads to the collapse of the gels under applied strains. It is also seen, using systematic electroacoustic measurements, that the gels exhibit considerable stability under gravity upto $C_{s}=300$ mM  (\autoref{Fig:TVI-salt}).  At $C_{s}>300$ mM, the gels become unstable due to   irreversible branch coagulation, with the gel network eventually collapsing under gravity. 

\indent  In conclusion, the present study elucidates the link between the bulk rheological and stability behaviors of   natural Na-montmorillonite gels and their underlying microscopic structures. Apart from the significant importance of this study for various rheological applications of Na-montmorillonite, we believe that the results presented here will facilitate our understanding of the dispersion behavior of other  smectite clays such as Laponite and Kaolinite in the presence of salt. Since  soil is composed of natural smectites, our results are also extremely useful in the understanding of many geophysical phenomena such as landslides, and the formation of quicksand and river deltas.

\section{Acknowledgment}
We thank Mr. A. Dhasan and Mr. K. M. Yatheendran for their  help with  cryo-SEM imaging.

\noindent  \textbf{FOOT NOTE:}\\

 \ddag Electronic supplementary information (ESI) available: SEM micrograph of Na-montmorillonite platelets (Fig. S1),  lateral size distribution of Na-montmorillonite platelets (Fig. S2), variation of pH of 5\% w/v Na-montmorillonite dispersion with salt concentration $C_{s}$ (Fig. S3),  experimental protocol of rheological measurements and variation of elastic modulus, $G'$, and viscous modulus, $G''$, during shear melting process  (Fig. S4),  evolution of $G'$ and $G''$ with age, $t_{w}$ (Fig. S5),    experimental setup for monitoring sedimentation stability of clay dispersions (Fig. S6), a magnified cryo-SEM micrograph of 5\% w/v Na-montmorillonite with $C_{s}=10$ mM (Fig. S7), analysis method of cryo-SEM images using ImageJ (Section A), a 2D projected binary form of a representative cryo-SEM micrograph (Fig. S8), pore size distribution of a 5\% w/v Na-montmorillonite gel with $C_{s}=20$ mM (Fig. S9),   and magnified cryo-SEM micrographs of 5\% w/v Na-montmorillonite samples with $C_{s}=$ 150 mM (Fig. S10), 300 mM (Fig. S11)  and 600 mM (Fig. S12). See DOI: *************

%%%%%%%%%%%%%%%%%%%%%%%%%%%%%%%%%%%%%%%%%%%%%%%%%%%%%%%%%%%%%%%%%%%%%%%%%%%%%%%%%%%%%%%%%%%%%%%%%%%%%%%%%%%%%%%%%%%%%%%%%%%%%%%%%%%%%%%%%%%%%%%%%%%%%%%%%%%%%%%%%%%%%%%%%%%%%%
 
\providecommand*{\mcitethebibliography}{\thebibliography}
\csname @ifundefined\endcsname{endmcitethebibliography}
{\let\endmcitethebibliography\endthebibliography}{}

%%%%%%%%%%%%%%%%%%%%%%%%%%%%%%%%%%%%%%%%%%%%%%%%%%%%%%%%%%%%%%%%%%%%%%%%%%%%%%%%%%%%%%%%%%%%%%%%%%%%%%%%%%%%%%%%%%%%%%%%%%%%%%%%%%%%%%%%%%%%%%%%%%%%%%%%%%%%%%%%%%%%%%%%%%%%%%

%\section{References}
%\bibliographystyle{rsc}
%\bibliography{References}

\begin{mcitethebibliography}{56}
\providecommand*{\natexlab}[1]{#1}
\providecommand*{\mciteSetBstSublistMode}[1]{}
\providecommand*{\mciteSetBstMaxWidthForm}[2]{}
\providecommand*{\mciteBstWouldAddEndPuncttrue}
  {\def\EndOfBibitem{\unskip.}}
\providecommand*{\mciteBstWouldAddEndPunctfalse}
  {\let\EndOfBibitem\relax}
\providecommand*{\mciteSetBstMidEndSepPunct}[3]{}
\providecommand*{\mciteSetBstSublistLabelBeginEnd}[3]{}
\providecommand*{\EndOfBibitem}{}
\mciteSetBstSublistMode{f}
\mciteSetBstMaxWidthForm{subitem}
{(\emph{\alph{mcitesubitemcount}})}
\mciteSetBstSublistLabelBeginEnd{\mcitemaxwidthsubitemform\space}
{\relax}{\relax}

\bibitem[Abend and Lagaly(2000)]{Abend.2000.1}
S.~Abend and G.~Lagaly, \emph{Applied Clay Science}, 2000, \textbf{16}, 201 --
  227\relax
\mciteBstWouldAddEndPuncttrue
\mciteSetBstMidEndSepPunct{\mcitedefaultmidpunct}
{\mcitedefaultendpunct}{\mcitedefaultseppunct}\relax
\EndOfBibitem
\bibitem[Jabbari-Farouji \emph{et~al.}(2008)Jabbari-Farouji, Tanaka, Wegdam,
  and Bonn]{Jabbari.PRE.2008}
S.~Jabbari-Farouji, H.~Tanaka, G.~H. Wegdam and D.~Bonn, \emph{Phys. Rev. E},
  2008, \textbf{78}, 061405\relax
\mciteBstWouldAddEndPuncttrue
\mciteSetBstMidEndSepPunct{\mcitedefaultmidpunct}
{\mcitedefaultendpunct}{\mcitedefaultseppunct}\relax
\EndOfBibitem
\bibitem[Bandyopadhyay \emph{et~al.}(2004)Bandyopadhyay, Liang, Yardimci,
  Sessoms, Borthwick, Mochrie, Harden, and Leheny]{Ranjini.PRL.2004}
R.~Bandyopadhyay, D.~Liang, H.~Yardimci, D.~A. Sessoms, M.~A. Borthwick,
  S.~G.~J. Mochrie, J.~L. Harden and R.~L. Leheny, \emph{Phys. Rev. Lett.},
  2004, \textbf{93}, 228302\relax
\mciteBstWouldAddEndPuncttrue
\mciteSetBstMidEndSepPunct{\mcitedefaultmidpunct}
{\mcitedefaultendpunct}{\mcitedefaultseppunct}\relax
\EndOfBibitem
\bibitem[Negi and Osuji(2010)]{Negi.JoR.2010}
A.~S. Negi and C.~O. Osuji, \emph{Journal of Rheology (1978-present)}, 2010,
  \textbf{54}, 943--958\relax
\mciteBstWouldAddEndPuncttrue
\mciteSetBstMidEndSepPunct{\mcitedefaultmidpunct}
{\mcitedefaultendpunct}{\mcitedefaultseppunct}\relax
\EndOfBibitem
\bibitem[Saha \emph{et~al.}(2014)Saha, Joshi, and
  Bandyopadhyay]{Debasish.SoftMatter.2014}
D.~Saha, Y.~M. Joshi and R.~Bandyopadhyay, \emph{Soft Matter}, 2014,
  \textbf{10}, 3292--3300\relax
\mciteBstWouldAddEndPuncttrue
\mciteSetBstMidEndSepPunct{\mcitedefaultmidpunct}
{\mcitedefaultendpunct}{\mcitedefaultseppunct}\relax
\EndOfBibitem
\bibitem[Tudisca \emph{et~al.}(2012)Tudisca, Ricci, Angelini, and
  Ruzicka]{tudisca.RSCAdv.2012}
V.~Tudisca, M.~A. Ricci, R.~Angelini and B.~Ruzicka, \emph{RSC Adv.}, 2012,
  \textbf{2}, 11111--11116\relax
\mciteBstWouldAddEndPuncttrue
\mciteSetBstMidEndSepPunct{\mcitedefaultmidpunct}
{\mcitedefaultendpunct}{\mcitedefaultseppunct}\relax
\EndOfBibitem
\bibitem[Saha \emph{et~al.}(2015)Saha, Bandyopadhyay, and
  Joshi]{debasish.Langmuir.2015}
D.~Saha, R.~Bandyopadhyay and Y.~M. Joshi, \emph{Langmuir}, 2015, \textbf{31},
  3012--3020\relax
\mciteBstWouldAddEndPuncttrue
\mciteSetBstMidEndSepPunct{\mcitedefaultmidpunct}
{\mcitedefaultendpunct}{\mcitedefaultseppunct}\relax
\EndOfBibitem
\bibitem[Fossum(2012)]{fossum.EPJ.2012}
J.~Fossum, \emph{The European Physical Journal Special Topics}, 2012,
  \textbf{204}, 41--56\relax
\mciteBstWouldAddEndPuncttrue
\mciteSetBstMidEndSepPunct{\mcitedefaultmidpunct}
{\mcitedefaultendpunct}{\mcitedefaultseppunct}\relax
\EndOfBibitem
\bibitem[Bailey \emph{et~al.}(2015)Bailey, Lekkerkerker, and
  Maitland]{Bailey.SoftMatter.2015}
L.~Bailey, H.~N.~W. Lekkerkerker and G.~C. Maitland, \emph{Soft Matter}, 2015,
  \textbf{11}, 222--236\relax
\mciteBstWouldAddEndPuncttrue
\mciteSetBstMidEndSepPunct{\mcitedefaultmidpunct}
{\mcitedefaultendpunct}{\mcitedefaultseppunct}\relax
\EndOfBibitem
\bibitem[Glotzer and Solomon(2007)]{Sharon.NatMat.2007}
S.~C. Glotzer and M.~J. Solomon, \emph{Nat Mater}, 2007, \textbf{6}, 557 --
  562\relax
\mciteBstWouldAddEndPuncttrue
\mciteSetBstMidEndSepPunct{\mcitedefaultmidpunct}
{\mcitedefaultendpunct}{\mcitedefaultseppunct}\relax
\EndOfBibitem
\bibitem[Ruzicka and Zaccarelli(2011)]{ruzicka.SoftMatter.2011}
B.~Ruzicka and E.~Zaccarelli, \emph{Soft Matter}, 2011, \textbf{7},
  1268--1286\relax
\mciteBstWouldAddEndPuncttrue
\mciteSetBstMidEndSepPunct{\mcitedefaultmidpunct}
{\mcitedefaultendpunct}{\mcitedefaultseppunct}\relax
\EndOfBibitem
\bibitem[Barbara \emph{et~al.}(2011)Barbara, Emanuela, Laura, Roberta, Michael,
  Abdellatif, Theyencheri, and Francesco]{ruzicka.NatureMaterial.2011}
R.~Barbara, Z.~Emanuela, Z.~Laura, A.~Roberta, S.~Michael, M.~Abdellatif,
  N.~Theyencheri and S.~Francesco, \emph{Nature Materials}, 2011, \textbf{10},
  56 -- 60\relax
\mciteBstWouldAddEndPuncttrue
\mciteSetBstMidEndSepPunct{\mcitedefaultmidpunct}
{\mcitedefaultendpunct}{\mcitedefaultseppunct}\relax
\EndOfBibitem
\bibitem[Paineau \emph{et~al.}(2013)Paineau, Philippe, Antonova, Bihannic,
  Davidson, Dozov, Gabriel, Imperor-Clerc, Levitz, Meneau, and
  Michot]{Paineau.LCR.2013}
E.~Paineau, A.~Philippe, K.~Antonova, I.~Bihannic, P.~Davidson, I.~Dozov,
  J.~Gabriel, M.~Imperor-Clerc, P.~Levitz, F.~Meneau and L.~Michot,
  \emph{Liquid Crystals Reviews}, 2013, \textbf{1}, 110--126\relax
\mciteBstWouldAddEndPuncttrue
\mciteSetBstMidEndSepPunct{\mcitedefaultmidpunct}
{\mcitedefaultendpunct}{\mcitedefaultseppunct}\relax
\EndOfBibitem
\bibitem[Uddin(2008)]{faheem.MMT-A.2008}
F.~Uddin, \emph{Metallurgical and Materials Transactions A}, 2008, \textbf{39},
  2804--2814\relax
\mciteBstWouldAddEndPuncttrue
\mciteSetBstMidEndSepPunct{\mcitedefaultmidpunct}
{\mcitedefaultendpunct}{\mcitedefaultseppunct}\relax
\EndOfBibitem
\bibitem[Murray(2000)]{Murray.2000.207}
H.~H. Murray, \emph{Applied Clay Science}, 2000, \textbf{17}, 207 -- 221\relax
\mciteBstWouldAddEndPuncttrue
\mciteSetBstMidEndSepPunct{\mcitedefaultmidpunct}
{\mcitedefaultendpunct}{\mcitedefaultseppunct}\relax
\EndOfBibitem
\bibitem[Viseras \emph{et~al.}(2007)Viseras, Aguzzi, Cerezo, and
  Lopez-Galindo]{viseras.ACSc.2007}
C.~Viseras, C.~Aguzzi, P.~Cerezo and A.~Lopez-Galindo, \emph{Applied Clay
  Science}, 2007, \textbf{36}, 37 -- 50\relax
\mciteBstWouldAddEndPuncttrue
\mciteSetBstMidEndSepPunct{\mcitedefaultmidpunct}
{\mcitedefaultendpunct}{\mcitedefaultseppunct}\relax
\EndOfBibitem
\bibitem[Liu \emph{et~al.}(2006)Liu, Zhu, Liu, Zhang, Sun, Chen, and
  Adler]{Liu.2006.1}
Y.~Liu, M.~Zhu, X.~Liu, W.~Zhang, B.~Sun, Y.~Chen and H.-J.~P. Adler,
  \emph{Polymer}, 2006, \textbf{47}, 1 -- 5\relax
\mciteBstWouldAddEndPuncttrue
\mciteSetBstMidEndSepPunct{\mcitedefaultmidpunct}
{\mcitedefaultendpunct}{\mcitedefaultseppunct}\relax
\EndOfBibitem
\bibitem[Bergaya and Lagaly(2013)]{Bergaya.2013.1}
F.~Bergaya and G.~Lagaly, in \emph{Handbook of Clay Science}, ed. F.~Bergaya
  and G.~Lagaly, Elsevier, 2013, vol.~5, pp. 1 -- 19\relax
\mciteBstWouldAddEndPuncttrue
\mciteSetBstMidEndSepPunct{\mcitedefaultmidpunct}
{\mcitedefaultendpunct}{\mcitedefaultseppunct}\relax
\EndOfBibitem
\bibitem[Secor and Radke(1985)]{Secor.JCIS.1985}
R.~Secor and C.~Radke, \emph{Journal of Colloid and Interface Science}, 1985,
  \textbf{103}, 237 -- 244\relax
\mciteBstWouldAddEndPuncttrue
\mciteSetBstMidEndSepPunct{\mcitedefaultmidpunct}
{\mcitedefaultendpunct}{\mcitedefaultseppunct}\relax
\EndOfBibitem
\bibitem[Yan and Eisenthal(1999)]{Yan.JPCB.1999}
E.~C.~Y. Yan and K.~B. Eisenthal, \emph{The Journal of Physical Chemistry B},
  1999, \textbf{103}, 6056--6060\relax
\mciteBstWouldAddEndPuncttrue
\mciteSetBstMidEndSepPunct{\mcitedefaultmidpunct}
{\mcitedefaultendpunct}{\mcitedefaultseppunct}\relax
\EndOfBibitem
\bibitem[Zhou \emph{et~al.}(2012)Zhou, Abdel-Fattah, and Keller]{Zhou.EST.2012}
D.~Zhou, A.~I. Abdel-Fattah and A.~A. Keller, \emph{Environmental Science \&
  Technology}, 2012, \textbf{46}, 7520--7526\relax
\mciteBstWouldAddEndPuncttrue
\mciteSetBstMidEndSepPunct{\mcitedefaultmidpunct}
{\mcitedefaultendpunct}{\mcitedefaultseppunct}\relax
\EndOfBibitem
\bibitem[Verwey and Overbeek(1948)]{Overbeek-book-1948}
E.~Verwey and J.~Overbeek, \emph{Theory of Stability of Lyophobic Colloids},
  Elsevier: Amsterdam, Netherlands, 1948\relax
\mciteBstWouldAddEndPuncttrue
\mciteSetBstMidEndSepPunct{\mcitedefaultmidpunct}
{\mcitedefaultendpunct}{\mcitedefaultseppunct}\relax
\EndOfBibitem
\bibitem[Tanaka \emph{et~al.}(2004)Tanaka, Meunier, and Bonn]{Tanaka.PRE.2004}
H.~Tanaka, J.~Meunier and D.~Bonn, \emph{Phys. Rev. E}, 2004, \textbf{69},
  031404\relax
\mciteBstWouldAddEndPuncttrue
\mciteSetBstMidEndSepPunct{\mcitedefaultmidpunct}
{\mcitedefaultendpunct}{\mcitedefaultseppunct}\relax
\EndOfBibitem
\bibitem[Norrish(1954)]{norrish.1954}
K.~Norrish, \emph{Discuss. Faraday Soc.}, 1954, \textbf{18}, 120--134\relax
\mciteBstWouldAddEndPuncttrue
\mciteSetBstMidEndSepPunct{\mcitedefaultmidpunct}
{\mcitedefaultendpunct}{\mcitedefaultseppunct}\relax
\EndOfBibitem
\bibitem[Bowles(1968)]{Bowles.Science.1968}
F.~A. Bowles, \emph{Science}, 1968, \textbf{159}, 1236--1237\relax
\mciteBstWouldAddEndPuncttrue
\mciteSetBstMidEndSepPunct{\mcitedefaultmidpunct}
{\mcitedefaultendpunct}{\mcitedefaultseppunct}\relax
\EndOfBibitem
\bibitem[van Olphen(1977)]{Olphen_book}
H.~van Olphen, \emph{John Wiley and Sons Inc.: New York}, 1977, \textbf{53},
  230--230\relax
\mciteBstWouldAddEndPuncttrue
\mciteSetBstMidEndSepPunct{\mcitedefaultmidpunct}
{\mcitedefaultendpunct}{\mcitedefaultseppunct}\relax
\EndOfBibitem
\bibitem[Michot \emph{et~al.}(2013)Michot, Bihannic, Thomas, Lartiges,
  Waldvogel, Caillet, Thieme, Funari, and Levitz]{Michot.Langmuir.2013}
L.~J. Michot, I.~Bihannic, F.~Thomas, B.~S. Lartiges, Y.~Waldvogel, C.~Caillet,
  J.~Thieme, S.~S. Funari and P.~Levitz, \emph{Langmuir}, 2013, \textbf{29},
  3500--3510\relax
\mciteBstWouldAddEndPuncttrue
\mciteSetBstMidEndSepPunct{\mcitedefaultmidpunct}
{\mcitedefaultendpunct}{\mcitedefaultseppunct}\relax
\EndOfBibitem
\bibitem[Duran \emph{et~al.}(2000)Duran, Ramos-Tejada, Arroyo, and
  Gonzalez-Caballero]{duran.JCIS.2000}
J.~Duran, M.~Ramos-Tejada, F.~Arroyo and F.~Gonzalez-Caballero, \emph{Journal
  of Colloid and Interface Science}, 2000, \textbf{229}, 107 -- 117\relax
\mciteBstWouldAddEndPuncttrue
\mciteSetBstMidEndSepPunct{\mcitedefaultmidpunct}
{\mcitedefaultendpunct}{\mcitedefaultseppunct}\relax
\EndOfBibitem
\bibitem[Lagaly and Ziesmer(2003)]{Lagaly.2003.1}
G.~Lagaly and S.~Ziesmer, \emph{Advances in Colloid and Interface Science},
  2003, \textbf{100?€“102}, 105 -- 128\relax
\mciteBstWouldAddEndPuncttrue
\mciteSetBstMidEndSepPunct{\mcitedefaultmidpunct}
{\mcitedefaultendpunct}{\mcitedefaultseppunct}\relax
\EndOfBibitem
\bibitem[Broughton and Squires(1935)]{broughton.JPC.1935}
G.~Broughton and L.~Squires, \emph{The Journal of Physical Chemistry}, 1935,
  \textbf{40}, 1041--1053\relax
\mciteBstWouldAddEndPuncttrue
\mciteSetBstMidEndSepPunct{\mcitedefaultmidpunct}
{\mcitedefaultendpunct}{\mcitedefaultseppunct}\relax
\EndOfBibitem
\bibitem[van Olphen(1964)]{vanOlphen.JCS.1964}
H.~van Olphen, \emph{Journal of Colloid Science}, 1964, \textbf{19}, 313 --
  322\relax
\mciteBstWouldAddEndPuncttrue
\mciteSetBstMidEndSepPunct{\mcitedefaultmidpunct}
{\mcitedefaultendpunct}{\mcitedefaultseppunct}\relax
\EndOfBibitem
\bibitem[Au and Leong(2013)]{Au.CSA.2013}
P.-I. Au and Y.-K. Leong, \emph{Colloids and Surfaces A: Physicochemical and
  Engineering Aspects}, 2013, \textbf{436}, 530 -- 541\relax
\mciteBstWouldAddEndPuncttrue
\mciteSetBstMidEndSepPunct{\mcitedefaultmidpunct}
{\mcitedefaultendpunct}{\mcitedefaultseppunct}\relax
\EndOfBibitem
\bibitem[Kimura \emph{et~al.}(2011)Kimura, Sakurai, Sugiyama, Tsuchida, Okubo,
  and Masuko]{kimura.RheoActa.2011}
H.~Kimura, M.~Sakurai, T.~Sugiyama, A.~Tsuchida, T.~Okubo and T.~Masuko,
  \emph{Rheologica Acta}, 2011, \textbf{50}, 159--168\relax
\mciteBstWouldAddEndPuncttrue
\mciteSetBstMidEndSepPunct{\mcitedefaultmidpunct}
{\mcitedefaultendpunct}{\mcitedefaultseppunct}\relax
\EndOfBibitem
\bibitem[Laxton and Berg(2006)]{Laxton.JCIS.2006}
P.~B. Laxton and J.~C. Berg, \emph{Journal of Colloid and Interface Science},
  2006, \textbf{296}, 749 -- 755\relax
\mciteBstWouldAddEndPuncttrue
\mciteSetBstMidEndSepPunct{\mcitedefaultmidpunct}
{\mcitedefaultendpunct}{\mcitedefaultseppunct}\relax
\EndOfBibitem
\bibitem[Dijkstra \emph{et~al.}(1995)Dijkstra, Hansen, and
  Madden]{dijkstra.PRL.1995}
M.~Dijkstra, J.~P. Hansen and P.~Madden, \emph{Phys. Rev. Lett.}, 1995,
  \textbf{75}, 2236--2239\relax
\mciteBstWouldAddEndPuncttrue
\mciteSetBstMidEndSepPunct{\mcitedefaultmidpunct}
{\mcitedefaultendpunct}{\mcitedefaultseppunct}\relax
\EndOfBibitem
\bibitem[Wierzchos \emph{et~al.}(1992)Wierzchos, Ascaso, Garcia-Gonzalez, and
  Kozak]{wierzchos.CCMin.1992}
J.~Wierzchos, C.~Ascaso, M.~T. Garcia-Gonzalez and E.~Kozak, \emph{Clays and
  Clay Minerals}, 1992, \textbf{40}, 230--236\relax
\mciteBstWouldAddEndPuncttrue
\mciteSetBstMidEndSepPunct{\mcitedefaultmidpunct}
{\mcitedefaultendpunct}{\mcitedefaultseppunct}\relax
\EndOfBibitem
\bibitem[Zbik \emph{et~al.}(2008)Zbik, Martens, Frost, Song, Chen, and
  Chen]{marek.Langmuir.2008}
M.~S. Zbik, W.~N. Martens, R.~L. Frost, Y.-F. Song, Y.-M. Chen and J.-H. Chen,
  \emph{Langmuir}, 2008, \textbf{24}, 8954--8958\relax
\mciteBstWouldAddEndPuncttrue
\mciteSetBstMidEndSepPunct{\mcitedefaultmidpunct}
{\mcitedefaultendpunct}{\mcitedefaultseppunct}\relax
\EndOfBibitem
\bibitem[Tanaka \emph{et~al.}(2005)Tanaka, Jabbari-Farouji, Meunier, and
  Bonn]{tanaka.PhysRevE.2005}
H.~Tanaka, S.~Jabbari-Farouji, J.~Meunier and D.~Bonn, \emph{Phys. Rev. E},
  2005, \textbf{71}, 021402\relax
\mciteBstWouldAddEndPuncttrue
\mciteSetBstMidEndSepPunct{\mcitedefaultmidpunct}
{\mcitedefaultendpunct}{\mcitedefaultseppunct}\relax
\EndOfBibitem
\bibitem[Angelini \emph{et~al.}(2014)Angelini, Zaccarelli, de~Melo~Marques,
  Sztucki, Fluerasu, Ruocco, and Ruzicka]{angelini.Naturecom.2014}
R.~Angelini, E.~Zaccarelli, F.~A. de~Melo~Marques, M.~Sztucki, A.~Fluerasu,
  G.~Ruocco and B.~Ruzicka, \emph{Nature communications}, 2014, \textbf{5},
  year\relax
\mciteBstWouldAddEndPuncttrue
\mciteSetBstMidEndSepPunct{\mcitedefaultmidpunct}
{\mcitedefaultendpunct}{\mcitedefaultseppunct}\relax
\EndOfBibitem
\bibitem[Thill \emph{et~al.}(2001)Thill, Moustier, Garnier, Estournel, Naudin,
  and Bottero]{Thill.CSR.2001}
A.~Thill, S.~Moustier, J.-M. Garnier, C.~Estournel, J.-J. Naudin and J.-Y.
  Bottero, \emph{Continental Shelf Research}, 2001, \textbf{21}, 2127 --
  2140\relax
\mciteBstWouldAddEndPuncttrue
\mciteSetBstMidEndSepPunct{\mcitedefaultmidpunct}
{\mcitedefaultendpunct}{\mcitedefaultseppunct}\relax
\EndOfBibitem
\bibitem[Khaldoun \emph{et~al.}(2005)Khaldoun, Eiser, Wegdam, and
  Bonn]{khaldoun.Nature.2005}
A.~Khaldoun, E.~Eiser, G.~Wegdam and D.~Bonn, \emph{Nature}, 2005,
  \textbf{437}, 635--635\relax
\mciteBstWouldAddEndPuncttrue
\mciteSetBstMidEndSepPunct{\mcitedefaultmidpunct}
{\mcitedefaultendpunct}{\mcitedefaultseppunct}\relax
\EndOfBibitem
\bibitem[nan((Last accessed on 22nd February 2015))]{nanocor-mmt}
\emph{Nanocor Inc., Technical data sheet},  G-105, (Last accessed on 22nd
  February 2015)\relax
\mciteBstWouldAddEndPuncttrue
\mciteSetBstMidEndSepPunct{\mcitedefaultmidpunct}
{\mcitedefaultendpunct}{\mcitedefaultseppunct}\relax
\EndOfBibitem
\bibitem[Olphen(1962)]{VanOlphen.1962.1}
H.~V. Olphen, \emph{Journal of Colloid Science}, 1962, \textbf{17}, 660 --
  667\relax
\mciteBstWouldAddEndPuncttrue
\mciteSetBstMidEndSepPunct{\mcitedefaultmidpunct}
{\mcitedefaultendpunct}{\mcitedefaultseppunct}\relax
\EndOfBibitem
\bibitem[Zheng and Zaoui(2011)]{Zheng.2011.80}
Y.~Zheng and A.~Zaoui, \emph{Solid State Ionics}, 2011, \textbf{203}, 80 --
  85\relax
\mciteBstWouldAddEndPuncttrue
\mciteSetBstMidEndSepPunct{\mcitedefaultmidpunct}
{\mcitedefaultendpunct}{\mcitedefaultseppunct}\relax
\EndOfBibitem
\bibitem[Ramsay \emph{et~al.}(1990)Ramsay, Swanton, and Bunce]{ramsay.1990.1}
J.~D.~F. Ramsay, S.~W. Swanton and J.~Bunce, \emph{J. Chem. Soc.{,} Faraday
  Trans.}, 1990, \textbf{86}, 3919--3926\relax
\mciteBstWouldAddEndPuncttrue
\mciteSetBstMidEndSepPunct{\mcitedefaultmidpunct}
{\mcitedefaultendpunct}{\mcitedefaultseppunct}\relax
\EndOfBibitem
\bibitem[Cadene \emph{et~al.}(2005)Cadene, Durand-Vidal, Turq, and
  Brendle]{Cadene.2005.1}
A.~Cadene, S.~Durand-Vidal, P.~Turq and J.~Brendle, \emph{Journal of Colloid
  and Interface Science}, 2005, \textbf{285}, 719 -- 730\relax
\mciteBstWouldAddEndPuncttrue
\mciteSetBstMidEndSepPunct{\mcitedefaultmidpunct}
{\mcitedefaultendpunct}{\mcitedefaultseppunct}\relax
\EndOfBibitem
\bibitem[Ali and Bandyopadhyay(2013)]{ali.samim.2013.1}
S.~Ali and R.~Bandyopadhyay, \emph{Langmuir}, 2013, \textbf{29},
  12663--12669\relax
\mciteBstWouldAddEndPuncttrue
\mciteSetBstMidEndSepPunct{\mcitedefaultmidpunct}
{\mcitedefaultendpunct}{\mcitedefaultseppunct}\relax
\EndOfBibitem
\bibitem[Ali and Bandyopadhyay(2015)]{ali.ACSs.2015}
S.~Ali and R.~Bandyopadhyay, \emph{Applied Clay Science}, 2015, \textbf{114},
  85--92\relax
\mciteBstWouldAddEndPuncttrue
\mciteSetBstMidEndSepPunct{\mcitedefaultmidpunct}
{\mcitedefaultendpunct}{\mcitedefaultseppunct}\relax
\EndOfBibitem
\bibitem[Dukhin and Goetz(2010)]{Dukhin-book-2nd}
A.~S. Dukhin and P.~J. Goetz, \emph{Characterization of Liquids, Nano- and
  Microparticulates, and Porous Bodies using Ultrasound}, Elsevier: New York,
  2010\relax
\mciteBstWouldAddEndPuncttrue
\mciteSetBstMidEndSepPunct{\mcitedefaultmidpunct}
{\mcitedefaultendpunct}{\mcitedefaultseppunct}\relax
\EndOfBibitem
\bibitem[Miyazaki \emph{et~al.}(2006)Miyazaki, Wyss, Weitz, and
  Reichman]{miyazaki.EPL.2006}
K.~Miyazaki, H.~M. Wyss, D.~A. Weitz and D.~R. Reichman, \emph{EPL (Europhysics
  Letters)}, 2006, \textbf{75}, 915\relax
\mciteBstWouldAddEndPuncttrue
\mciteSetBstMidEndSepPunct{\mcitedefaultmidpunct}
{\mcitedefaultendpunct}{\mcitedefaultseppunct}\relax
\EndOfBibitem
\bibitem[Laurati \emph{et~al.}(2011)Laurati, Egelhaaf, and
  Petekidis]{laurati.JOR.2011}
M.~Laurati, S.~U. Egelhaaf and G.~Petekidis, \emph{Journal of Rheology
  (1978-present)}, 2011, \textbf{55}, 673--706\relax
\mciteBstWouldAddEndPuncttrue
\mciteSetBstMidEndSepPunct{\mcitedefaultmidpunct}
{\mcitedefaultendpunct}{\mcitedefaultseppunct}\relax
\EndOfBibitem
\bibitem[Jonsson \emph{et~al.}(2008)Jonsson, Labbez, and
  Cabane]{jonsson.Langmuir.2008}
B.~Jonsson, C.~Labbez and B.~Cabane, \emph{Langmuir}, 2008, \textbf{24},
  11406--11413\relax
\mciteBstWouldAddEndPuncttrue
\mciteSetBstMidEndSepPunct{\mcitedefaultmidpunct}
{\mcitedefaultendpunct}{\mcitedefaultseppunct}\relax
\EndOfBibitem
\bibitem[Delhorme \emph{et~al.}(2012)Delhorme, Jonsson, and
  Labbez]{delhorme.SoftMatter.2012}
M.~Delhorme, B.~Jonsson and C.~Labbez, \emph{Soft Matter}, 2012, \textbf{8},
  9691--9704\relax
\mciteBstWouldAddEndPuncttrue
\mciteSetBstMidEndSepPunct{\mcitedefaultmidpunct}
{\mcitedefaultendpunct}{\mcitedefaultseppunct}\relax
\EndOfBibitem
\bibitem[Weiss and Frank(1961)]{weiss.Naturforsch.1961}
A.~Weiss and R.~Frank, \emph{Naturforsch}, 1961, \textbf{16}, 141--142\relax
\mciteBstWouldAddEndPuncttrue
\mciteSetBstMidEndSepPunct{\mcitedefaultmidpunct}
{\mcitedefaultendpunct}{\mcitedefaultseppunct}\relax
\EndOfBibitem
\bibitem[Permien and Lagaly(1994)]{permien.ACSc.1994}
T.~Permien and G.~Lagaly, \emph{Applied Clay Science}, 1994, \textbf{9}, 251 --
  263\relax
\mciteBstWouldAddEndPuncttrue
\mciteSetBstMidEndSepPunct{\mcitedefaultmidpunct}
{\mcitedefaultendpunct}{\mcitedefaultseppunct}\relax
\EndOfBibitem
\bibitem[Luckham and Rossi(1999)]{Luckham.ACIS.1999}
P.~F. Luckham and S.~Rossi, \emph{Advances in Colloid and Interface Science},
  1999, \textbf{82}, 43 -- 92\relax
\mciteBstWouldAddEndPuncttrue
\mciteSetBstMidEndSepPunct{\mcitedefaultmidpunct}
{\mcitedefaultendpunct}{\mcitedefaultseppunct}\relax
\EndOfBibitem
\end{mcitethebibliography}
%\bibliographystyle{unsrt}

 %\begin{thebibliography}{99}

 %\end{thebibliography}

\end{document}